\documentclass[preprint, showpacs,preprintnumbers,amsmath,amssymb,nofootinbib]{revtex4}
\usepackage{amssymb}

\usepackage{graphicx}
\usepackage{dcolumn}
\usepackage{bm}

\newcommand{\bea}{\begin{eqnarray}}
\newcommand{\eea}{\end{eqnarray}}

\begin{document}
\title{Reconstructing the properties of dark energy from recent observations}
\author{  Puxun Wu\;$^{a}$ and  Hongwei Yu\;$^{b,}$\footnote{corresponding
author}}

\address
{$^a$Institute of  Math-Physics and School  of Sciences, Central
South University of Forestry and Technology, Changsha, Hunan 410004,
China
 \\
$^b$Department of Physics and Institute of  Physics,\\ Hunan Normal
University, Changsha, Hunan 410081, China
 }

\begin{abstract}
We explore the properties of dark energy from recent observational
data, including the Gold Sne Ia, the baryonic acoustic oscillation
peak from SDSS, the CMB shift parameter from WMAP3, the X-ray gas
mass fraction in cluster and the Hubble parameter versus redshift.
The $\Lambda CDM$ model with curvature and two parameterized dark
energy models are studied. For the $\Lambda CDM$ model, we find that
the flat universe is consistent with observations at the $1\sigma$
 confidence level and a closed universe is
slightly favored by these data. For two parameterized dark energy
models, with the prior given on the present matter density,
$\Omega_{m0}$,  with $\Omega_{m0}=0.24$, $\Omega_{m0}=0.28$ and
$\Omega_{m0}=0.32$, our result seems to suggest that the trend of
$\Omega_{m0}$ dependence for an evolving dark energy from a
combination of the observational data sets is model-dependent.

\end{abstract}

\pacs{ 98.80.-k, 98.80.Es }

 \maketitle

\section{Introduction}
The present cosmic accelerating expansion has been confirmed by
various observations, including the Type Ia Supernovae (Sne
Ia)~\cite{Perlmutter1999, Riess1998, Riess2004, Riess2006,
Astier2006, Wood2007}, CMB \cite{Balbi2000, de2000,
Jaffe2001,Spergel2003, Spergel2006} and large scale structure (LSS)
\cite{Peacock2001, Eisenstein2005}, etc. In order to explain this
observed phenomenon,  it is usually assumed that there exists, in
the universe,  an exotic energy component with negative pressure,
named dark energy (see  \cite{Padmanabhan2006, Copeland2006,
Sahni2006, Perivolaropoulos2006}  for recent reviews), which
presumably began to dominate the evolution of the universe only
recently. The simplest candidate of dark energy is the cosmological
constant $\Lambda$~\cite{Weinberg1989, Sahni2000, peebles2003,
Padmanabhan2003}. It fits the observational data very well, but at
the same time, it also encounters two problems, i.e.,  the
cosmological constant problem (why is the inferred value of
cosmological constant so tiny ($120$ orders of magnitude lower)
compared to the typical vacuum energy values predicted by particle
physics?) and the coincidence problem (why is its energy density
comparable to the matter density right now?). Therefore,  some
dynamical scalar fields, such as quintessence~\cite{Wetterich1988,
Ratra1988, Caldwell1998}, phantom~\cite{Cald} and
quintom~\cite{Quintom}, etc, are suggested as alternative candidates
of dark energy.  One of the features of these scalar field models is
that their equations of state parameter, $w$,  which embodies both
gravitational and evolutionary  properties of dark energy, is
evolving  with the cosmic expansion.

On the other hand, the growing number of dark energy models has
prompted people to adopt a complementary approach, which assumes an
arbitrary parametrization for the equation of state $w(z)$ in a
model-independent way and aims to reconstruct the properties of dark
energy directly from observations. Currently, there are many model
independent parameterizations (see for example,
~\cite{line1,line2,Jassal2005,Sahni2003, Sahni2004, Wetterich2004}).
In general, using these parameterizations and the observational
data, one can determine the present value of $w$ and whether it
evolves as the universe expands, in particular, whether the phantom
divide line (PDL) is crossed. In this regard,   Nesseris and
Perivolaropoulos \cite{Nesseris2006} has used the
Chevallier-Polarski-Linder parametrization $ w(z) = w_0 + w_1 z/(1 +
z)$~\cite{line2} to explore the properties of dark energy with some
observational data (including new Gold Sne Ia, SNLS Sne Ia, CMB,
BAO, the cluster baryon gas mass fraction(CBF)  and 2dF galaxy
redshift survey(2dFGRS) ) and found that the Gold data set mildly
favors dynamically evolving dark energy with the crossing of the PDL
while the SNLS does not,  and the combination of CMB+BAO+CBF+2dFGRS
 mildly favors the crossing of PDL only for low values of
 $\Omega_{m0}$ ( $\Omega_{m0}\le 0.25$)
prior considered and with  a higher prior matter density the
evolving features of dark energy becomes weaker and weaker. Similar
trend of  $\Omega_{m0}$ dependence was found using the
model~\cite{Alam2006},
$w(z)=\frac{1+z}{3}\frac{A_1+2A_2(1+z)}{\Omega_{DE}}-1$, with the
CMB and BAO.  However, constraints from a combination of the
supernovae and other observational data has not been analyzed in
Ref.~\cite{Nesseris2006}, and although that of the Sne and CMB+BAO
was examined in Ref.~\cite{Alam2006}, but the marginalization was
considered only for $\Omega_{m0}=0.28\pm 0.03$ prior. Therefore, it
remains interesting to see what happens to the conclusions reached
in Refs.~\cite{Nesseris2006,Alam2006}, when the combination of all
observational data is analyzed for different $\Omega_{m0}$ prior
considered. The present paper aims to fill the gap. We discuss the
constraints from the combination of different observational
datasets. Besides the data sets of Sne Ia, BAO and CMB, in our
analysis we add the datasets of the X-ray gas mass fraction in
cluster and the Hubble parameter versus redshift. Firstly the
$\Lambda CDM$ model with curvature is discussed. Then, two
parameterized  dark energy models: $ w(z) = w_0 + w_1 z/(1 + z)$ and
$w(z)=\frac{1+z}{3}\frac{A_1+2A_2(1+z)}{\Omega_{DE}}-1$, are studied
to see if the properties of dark energy thus reconstructed are
model-independent.

\section{The observational data }
\subsection{The Gold Sne Ia data}
The Sne Ia data considered in this paper is the 182 Gold set. This
set was released by Riess et al.~\cite{Riess2006} with a consistent
and robust manner. It consists of 119 previously published data
points~\cite{Riess2004}, 16 points discovered recently by the Hubble
Space Telescope (HST) and 47 points from the first year release of
the SNLS dataset~\cite{Astier2006}.  For these Sne Ia, the data is
released with the form of distance modulus $\mu$, which is relative
with the luminosity distance $d^L$ through
\begin{eqnarray}
\mu(z,H_0,p_j)=5\log_{10}[d^L(z,p_j)]+\mathcal{M}\;.
\end{eqnarray}
Here  for a flat universe $d^L=(1+z)\int_0^zdz'/E(z',p_j)$ with
$E^2(z)=H^2(z)/H^2_0=\Omega_{m0}+(1-\Omega_{m0})exp[3\int_0^z\frac{dz'}{1+z'}(1+w(z'))]$,
$\mathcal{M}=M-5log_{10}(H_0)$, $H_0=100h$, $M$ is the absolute
magnitude of the object and $p_j$  denote  the model parameters of
dark energy. The constraints on the dark energy models from Sne Ia
data can be obtained by the maximum likelihood method, so the best
fit values for model parameters is determined by minimizing
 \bea
\chi^2_{Sne}(H_0,p_j)=\Sigma_i\frac{[\mu(z_i,H_0,p_j)-\mu_{obs,i}]^2}{\sigma^2_{\mu_{obs,i}}}\eea

\subsection{The baryonic acoustic oscillation peak}
From the large scale correlation function of luminous red
galaxy in the Sloan Digital Sky Survey (SDSS), Eisenstein et
al.~\cite{Eisenstein2005} found a baryonic acoustic oscillation
peak, which is consistent with the prediction from the acoustic
oscillation in the primordial baryon-photon plasma at the
recombination. Thus this peak remarkably confirms the Big Bang
cosmology. Meanwhile it also provides a ruler to constrain the dark
energy models, which can be used usually by a dimensionless
parameter $A$,
\begin{eqnarray}
A(p_j)=\frac{\sqrt{\Omega_{m0}}}{E(z_1,p_j)^{1/3}}\bigg[\frac{1}{z_1}
  \int_0^{z_1}\frac{dz}{E(z,p_j)}\bigg]^{2/3}\;,
\end{eqnarray}
for a flat universe,  where $z_1 = 0.35$ and $A$ is measured to be
$A = 0.469\pm 0.017$. The parameter $A$ is  model-independent and
clearly independent of the value of $h$ too.
 By minimizing
\bea \chi^2_{BAO}(p_j)=\frac{[A(p_j)-0.469]^2}{0.017^2}\eea we can
obtain the constraints from BAO.

\subsection{The CMB shift parameter}
For the CMB data, we use the shift parameter $R$  to research the
properties of  dark energy, which can be expressed as~\cite{shiftR}
 \begin{eqnarray}
 R(p_j)=\sqrt{\Omega_{m0}}\int_0^{z_r}\frac{dz}{E(z,p_j)}\;,
\end{eqnarray}
for a flat universe, where $z_r =
1048[1+0.00124(\Omega_bh^2)^{-0.738}][1+g_1(\Omega_{m0}h^2)^{g_2}]$,
$g_1=0.078(\Omega_bh^2)^{-0.238}[1+39.5(\Omega_bh^2)^{0.763}]^{-1}$
and $g_2=0.56[1+21.1(\Omega_bh^2)^{1.81}]^{-1}$ \cite{Hu1996}. To
calculate $z_r$, we let $\Omega_bh^2=0.024$. The results of
three-year WMAP data~\cite{Spergel2006} give $R = 1.70 \pm
0.03$~\cite{Wang}. Let us note that the quantity $R$ from CMB
measurement is dependent on the value of $\Omega_{m0}h^2$, thus when
using this parameter it is required to marginalize over
$\Omega_{m0}h^2$ or assign some specific value to $h$. In our
discussion we give a prior value $h=0.72$.  We then place
constraints on cosmological models using this shift parameter by
minimizing  \bea
\chi^2_{CMB}(p_j)=\frac{[R(p_j)-1.70]^2}{0.03^2}\;.\eea

\subsection{The baryon gas mass fraction of galaxy cluster}
Under the basic assumption: the baryon gas mass fraction in cluster
is constant, a comparison of the gas mass fraction of galaxy
clusters, $f_{gas} = M_{gas}/M_{tot}$, can be used to constrain the
cosmological models. Following Allen et
al.~\cite{Allen2002,Allen2004} we fit the $f_{gas}$ data to a model
described by
 \bea f_{gas}^{mod}(z,p_j)=\frac{b\Omega_b (2h)^{3/2}}{(1+0.19h^{1/2})\Omega_{m0}
}\bigg[\frac{d_A^{SCDM}}{d_A^{mod}(z,p_j)}\bigg]^{3/2}\;,\eea where
$b$ is a parameter motivated by gas dynamical simulations,
$d_A=d^L/(1+z)^2$ is the angular diameter distance,  $d_A^{SCDM}$ is
the angular diameter distance corresponding to the standard cold
dark matter (SCDM) universe ($\Omega_{m0}=1$ for a flat universe).
Following Nesseris and Perivolaropoulos~\cite{Nesseris2006}, we
define $\lambda=\frac{b\Omega_b
(2h)^{3/2}}{(1+0.19h^{1/2})\Omega_{m0}}$ and treat  it as a nuisance
parameter. Using the theoretical method given in~\cite{Nesseris2006}
to  marginalize over $\lambda$, we can obtain the constraints by
minimizing \bea \chi^2_{f_{gas}}(p_j)=C-\frac{B^2}{A}\;,\eea where
 $A=\sum_i
\big(\frac{\widetilde{f}_{gas}^{mod}}{\sigma_{f_{gas,i}}}\big)^2$,
$B=\sum_i \frac{\widetilde{f}_{gas}^{mod}f_{gas,i}^{obs}
}{\sigma^2_{f_{gas,i}}}$, $C=\sum_i \big(\frac{f_{gas,i}^{obs}
}{\sigma^2_{f_{gas,i}}}\big)^2$, and
$\widetilde{f}_{gas}=\big[\frac{d_A^{SCDM}}{d_A^{mod}(z,p_j)}\big]^{3/2}
$.  Here 26 cluster data points given in Ref.~\cite{Allen2004} are
used.

\subsection{The Hubble parameter data}
Based on differential ages of passively evolving galaxies determined
from the Gemini Deep Deep Survey~\cite{Abraham2004} and archival
data~\cite{archival} at redshift $0\lesssim z\lesssim 1.8$,  Simon
et al.~\cite{Simon} obtained $9$ data points of $H(z)$ at redshift
$z_i$, which can be used to test the cosmological models by
minimizing \bea
\chi^2_H(H_0,p_j)=\Sigma_{i}\frac{[H_{obs}(z_i)-H_{th}(z_i,H_0,p_j)]^2}{\sigma_{Hi}^2}\;,\eea
Recently these 9 Hubble parameter data points have been studied
extensively by many authors~\cite{Simon, Wei2007}, however it doest
not provide a tight constraint on dark energy models.

Thus the constraints on cosmological models from a combination of
above discussed observational datasets  can be obtained by
minimizing
 \begin{eqnarray}
 \chi^2(H_0,p_j)=\chi^2_{Sne}(H_0,p_j)+\chi^2_{BAO}(p_j)+\chi^2_{CMB}(p_j)+\chi^2_{fgas}(p_j)+\chi^2_H(H_0,p_j)\;.
 \end{eqnarray}
Since we are interested in the model parameters,  $H_0$ becomes a
nuisance parameter and is marginalized   by a theoretical method
given in Ref.~\cite{Nesseris2004} in calculating  $\chi^2_{Sne}$ and
$\chi^2_{H}$.
\section{results}
The $\Lambda CDM$ model with curvature is firstly discussed with the
observational data. The results are shown in Fig.~1 for a
combination of the above discussed observational data sets. At a
$95.4\%$ confidence level we obtain
$\Omega_{m0}=0.29_{-0.04}^{+0.04}$ and
$\Omega_{k0}=-0.016_{-0.029}^{+0.030}$ with $\chi^2=196.8$. It is
easy to see that a spatially flat universe is consistent with the
observations at a $68\%$ confidence level and a closed universe is
somewhat favored by these data sets.

\begin{table}[!h]
\tabcolsep 0pt \caption{The best-fit data of Mod1 with prior value
$\Omega_{m0}$. In the Table G+C+B, G+C+B+f and G+C+B+f+H represent
the Gold+BAO+CMB, Gold+BAO+CMB+fgas and  Gold+BAO+CMB+fgas+H(z)
respectively. } \vspace*{-12pt}
\begin{center}
\def\temptablewidth{1\textwidth}
{\rule{\temptablewidth}{1pt}}
\begin{tabular*}{\temptablewidth}{@{\extracolsep{\fill}}|c|ccc|ccc|ccc|}
         & & $\Omega_{m0}=0.24$ &
         & & $\Omega_{m0}=0.28$ &
         & & $\Omega_{m0}=0.32$ & \\ \hline
         & $w_0$ & $w_1$           & $\chi^2_{Min}$
         & $w_0$ & $w_1$           & $\chi^2_{Min}$
         & $w_0$ & $w_1$           & $\chi^2_{Min}$ \\   \hline
  Gold  & $-1.28^{+0.59}_{-0.63}$ &  $2.64^{+3.08}_{-3.14}$  &156.5
       & $-1.38^{+0.67}_{-0.72}$ & $2.75^{+3.52}_{-3.73}$  &156.5
        & $-1.49^{+0.79}_{-0.85}$ & $2.81^{+4.13}_{-4.62}$  &156.6\\
  Gold+BAO & $-1.45^{+0.52}_{-0.55}$   &  $3.32^{+2.79}_{-2.63}$& 158.1
           & $-1.30^{+0.53}_{-0.59}$   &  $2.43^{+3.07}_{-3.40}$& 156.8
           & $-1.11^{+0.65}_{-0.64}$   &  $1.25^{+3.50}_{-4.28}$& 160.7  \\
  CMB+BAO  & $-1.47^{+0.85}_{-0.69}$ &  $1.48^{+1.51}_{-5.18}$  &0.001
           & $-1.01^{+0.94}_{-0.71}$ &  $0.61^{+1.40}_{-6.08}$  &0.001
           & $-0.60^{+1.05}_{-0.71}$ &  $-0.37^{+1.82}_{-6.75}$  &0.001  \\
  Gold+CMB  & $-0.90^{+0.42}_{-0.37}$   &  $0.27^{+1.00}_{-1.98}$& 161.8
            & $-1.05^{+0.44}_{-0.40}$   &  $0.81^{+0.81}_{-2.08}$& 158.6
            & $-1.22^{+0.43}_{-0.56}$   &  $1.28^{+0.73}_{-1.78}$& 157.4  \\
  G+C+B & $-1.04^{+0.35}_{-0.31}$   &  $0.63^{+0.79}_{-1.62}$& 165.1
        & $-1.05^{+0.39}_{-0.34}$   &  $0.79^{+0.74}_{-1.91}$& 158.6
        & $-1.06^{+0.44}_{-0.36}$   &  $0.96^{+0.68}_{-2.24}$& 160.8   \\
  G+C+B+f & $-1.07^{+0.35}_{-0.30}$   &  $0.68^{+0.76}_{-1.71}$& 191.1
          & $-1.10^{+0.40}_{-0.32}$   &  $0.88^{+0.71}_{-2.08}$& 185.9
          & $-1.13^{+0.47}_{-0.35}$   &  $1.10^{+0.62}_{-2.54}$& 189.6 \\
  G+C+B
  +f+H & $-1.07^{+0.30}_{-0.30}$   &  $0.71^{+0.73}_{-1.42}$& 200.2
      & $-1.09^{+0.36}_{-0.32}$   &  $0.85^{+0.70}_{-1.84}$& 194.9
      & $-1.12^{+0.47}_{-0.35}$   &  $1.06^{+0.65}_{-2.54}$& 199.1
       \end{tabular*}
       {\rule{\temptablewidth}{1pt}}
       \end{center}
       \end{table}

\begin{table}[!h]
\tabcolsep 0pt \caption{The best-fit data of Mod2  with prior value
$\Omega_{m0}$.  In the Table G+C+B, G+C+B+f and G+C+B+f+H represent
the Gold+BAO+CMB, Gold+BAO+CMB+fgas and  Gold+BAO+CMB+fgas+H(z)
respectively. } \vspace*{-12pt}
\begin{center}
\def\temptablewidth{1\textwidth}
{\rule{\temptablewidth}{1pt}}
\begin{tabular*}{\temptablewidth}{@{\extracolsep{\fill}}|c|ccc|ccc|ccc|}
         & &$\Omega_{m0}=0.24$ & & &$\Omega_{m0}=0.28$  & & &  $\Omega_{m0}=0.32$ & \\
         & $A_1$ & $A_2$ & $\chi^2_{Min}$   & $A_1$ & $A_2$           & $\chi^2_{Min}$  & $A_1$ & $A_2$           & $\chi^2_{Min}$\\   \hline
  Gold    & $-3.61^{+4.96}_{-5.52}$ &  $1.60^{+2.25}_{-1.95}$  &156.5
          & $-3.39^{+4.97}_{-5.53}$ &  $1.44^{+2.25}_{-1.95}$  &156.5
          & $-3.17^{+4.98}_{-5.54}$ &  $1.27^{+2.26}_{-1.95}$  &156.5  \\
  Gold+BAO & $-4.69^{+4.55}_{-5.17}$   &  $2.00^{+2.14}_{-1.83}$& 158.3
           & $-3.01^{+4.48}_{-5.09}$   &  $1.30^{+1.24}_{-1.13}$& 156.7
           & $-1.36^{+4.40}_{-5.02}$   &  $0.62^{+2.07}_{-1.75}$& 160.6  \\
  CMB+BAO  & $-1.49^{+2.52}_{-2.25}$ &  $0.41^{+0.72}_{-0.71}$  &0.001
           & $-0.21^{+2.90}_{-2.58}$ &  $0.14^{+0.81}_{-0.79}$  &0.001
           & $1.08^{+3.29}_{-2.91}$ &  $-0.12^{+0.89}_{-0.88}$  &0.001  \\
  Gold+CMB & $0.67^{+1.60}_{-1.70}$   &  $-0.11^{+0.57}_{-0.50}$& 161.6
           & $-0.33^{+1.82}_{-1.96}$   &  $0.20^{+0.68}_{-0.59}$& 159.1
           & $-1.38^{+2.02}_{-2.19}$   &  $0.55^{+0.78}_{-0.69}$& 157.4  \\
  G+C+B &  $0.11^{+1.49}_{-1.60}$   &  $0.04^{+0.56}_{-0.48}$& 166.1
               &  $-0.35^{+1.65}_{-1.77}$   &  $0.21^{+0.63}_{-0.55}$& 159.2
           &  $-0.70^{+1.79}_{-1.93}$   &  $0.35^{+0.71}_{-0.62}$& 160.7   \\
  G+C+B+f & $-0.05^{+1.42}_{-1.52}$   &  $0.07^{+0.54}_{-0.46}$& 192.1
              & $-0.53^{+1.58}_{-1.69}$   &  $0.25^{+0.61}_{-0.54}$& 186.3
              & $-0.91^{+1.71}_{-1.85}$   &  $0.40^{+0.68}_{-0.60}$& 189.2 \\
  G+C+B+f+H &  $-0.16^{+1.44}_{-1.42}$   &  $0.10^{+0.52}_{-0.45}$& 201.5
                         &  $-0.47^{+1.54}_{-1.60}$   &  $0.23^{+0.57}_{-0.51}$& 195.3
                         &  $-0.73^{+1.62}_{-1.71}$   &  $0.33^{+0.63}_{-0.56}$& 198.8
       \end{tabular*}
       {\rule{\temptablewidth}{1pt}}
       \end{center}
       \end{table}

Then we study, in the spatially flat case, the observational
constraints on the following two-parameter models considered in
Ref.~\cite{Nesseris2006} and ~\cite{Alam2006} respectively.
\begin{eqnarray}
Mod1:\quad w(z) = w_0 + w_1 \frac{z}{1 + z}\;,\end{eqnarray}
 \begin{eqnarray} Mod2:\quad w(z)=\frac{1+z}{3}\frac{A_1+2A_2(1+z)}{\Omega_{DE}}-1\;,\end{eqnarray}
where $\Omega_{DE}=A_1(1+z)+A_2(1+z)^2+1-\Omega_{m0}-A_1-A_2$
~\cite{Alam2006}. In order to find out whether the constraints from
the observations are dependent on the choice of the value of
$\Omega_{m0}$, we give three prior values of $\Omega_{m0}$ with
$\Omega_{m0}=0.24$, $\Omega_{m0}=0.28$ and $\Omega_{m0}=0.32$.
Tables 1 and 2  display constraints on model parameters with $95\%$
confidence level. From the best fit values given in the Tables one
can see that the BAO + CMB alone indicates that the evolving
features of dark energy become weaker and weaker with a higher prior
matter density, which is the same as obtained in
Refs.~\cite{Nesseris2006, Alam2006}, however the trend is reversed
for the Gold + CMB, Gold+CMB+BAO, Gold+CMB+BAO+fgas and
Gold+CMB+BAO+fgas+H(z). Meanwhile we find the effects of adding the
f-gas and Hubble parameter data sets are not very significant,
suggesting that the model parameters are strongly constrained by
CMB+BAO+Gold. In Fig.~2 we show the constraints on model parameters
from all the data sets. The upper and down panels show the
constraints on Mod1 and Mod2 respectively. In Fig.~2 the red dot
denotes the flat $\Lambda CDM$ model. This figure clearly shows that
$\Lambda CDM$ model is consistent with the observations at the
$95\%$ confidence level. In addition one can find that the best fit
of model parameters is closest to the $\Lambda CDM$ when
$\Omega_{m0}=0.24$, which can also be seen from the Tables.

 In Fig.~3   we plot the evolutionary curves of
$w(z)$ for these dark energy models with different prior values over
$\Omega_{m0}$. The upper and down panels show the results of Mod1
and Mod2 respectively. In Fig.~3 the solid lines show the evolution
of $w(z)$ with the best fit values for the model parameters, and the
dotted lines are for $1 \sigma$ and $2 \sigma$ error bars. For Mod1
the best fit curves show that the combination of the data sets
considered in this paper favors an evolving dark energy and a
crossing of the phantom divide line in the near past, and suggests
that the present value of $w$ is very likely less than $-1$.
Remarkably, these conclusions are almost insensitive to the chosen
value of matter density, in a clear contrast to those obtained in
~\cite{Nesseris2006} where
 it was found that, when
the Sne data are not combined with other observational ones,  the
evolving property becomes weaker and weaker with higher matter
density
 and the phantom divide may not be crossed with the increasing of matter
 density. With
this being said, it should be pointed out, however, that higher
values of $\Omega_{m0}$ lead to larger errors, especially at the
$2\sigma$ level, as can be seen from the Tables. The results for
Mod2 are shown in the down panels of Fig.~3. Unlike Mod1, the best
fit curves show that the properties reconstructed depend on the
chosen value of matter density. For $\Omega_{m0}=0.24$, a very
mildly evolving dark energy is obtained, but with the increase of
$\Omega_{m0}$ prior considered, the evolving feature of dark energy
becomes more evident. This trend is just the opposite to that found
in Ref.~\cite{Alam2006} for just BAO+CMB data. These discrepancies
can also be found in Tables 1 and 2. However at the $2\sigma$
confidence level for Mod1 and Mod2 the cosmological constant cannot
be ruled out.  In addition, we also find that the stringent
constraint on $w(z)$ happens around redshift $z\sim 0.5$, which is
consistent with that obtained in Refs.~\cite{Sahni2004, Gong}.

Finally, the evolution  of the decelerating parameter $q$ with the
redshift is studied for these two parameterized dark energy models
with $\Omega_{m0}=0.28$. The results are shown in Fig.~4 with the
left and right panel corresponding to  the results of Mod1 and Mod2
respectively. In this figure the dashed lines are the results of
$\Lambda CDM$ model with $\Omega_{m0}=0.28$, the solid lines show
the evolution of  $q(z)$ when $\Omega_{m0}=0.28$, the model
parameters are the best fit values, and the dotted lines are for $1
\sigma$ errors. The figure shows that the present value of the
deceleration parameter $q_0$ is less than zero,  indicating that the
cosmic is undergoing an accelerating expansion, but for different
models the value of  $q_0$ is different.

\section{conclusion}
In this paper, we have reconstructed the properties of dark energy
from recent observational data, including the Gold Sne Ia,  the
baryonic acoustic oscillation peak from SDSS, the CMB shift
parameter, the X-ray gas mass fraction in clusters and the Hubble
parameter data. The $\Lambda CDM$ model with curvature and two
parameterized dark energy models are discussed. We find that a
spatially flat universe is allowed by these data sets at the $68\%$
confidence level, and a closed universe is slightly favored by the
observations. For two parameterized dark energy models, we give the
priors on $\Omega_{m0}$ with $\Omega_{m0}=0.24$, $\Omega_{m0}=0.28$
and $\Omega_{m0}=0.32$. For the spatially flat case, the constraints
on model parameters and the evolutions of $w(z)$ and $q(z)$ are
studied. The Gold + CMB +BAO give the strong constraints on model
parameters. If Mod1 parametrization is used, the best fit curves in
Fig.~3 show  that the combination of the data sets considered in
this paper favors an evolving dark energy, a crossing of the phantom
divide line in the near past, and the present value of $w$ being
very likely less than $-1$. Remarkably, these conclusions are almost
insensitive to the chosen value of matter density, in a sharp
contrast to those obtained in~\cite{Nesseris2006} where the Sne data
are not combined with other observational ones.  However, the best
fit curves in Fig.~3 indicate that the properties of dark energy
reconstructed using Mod2 parametrization depend on the chosen value
of matter density. For $\Omega_{m0}=0.24$, a very mildly evolving
dark energy is obtained, but with the increase of $\Omega_{m0}$
prior considered, the evolving feature of dark energy becomes more
evident. This trend is just the opposite to that found in Ref.
\cite{Alam2006} for just BAO+CMB data.  Therefore, our result seems
to suggest that the trend of $\Omega_{m0}$ dependence for an
evolving dark energy is model-dependent. It should be noted,
however, that at the $2\sigma$ confidence level, the cosmological
constant are allowed for both Mod1 and Mod2.

\begin{acknowledgments}
This work  was supported in part by the National Natural Science
Foundation of China  under Grants No. 10575035,  10775050 and 
10705055, the Program for NCET under Grant No. 04-0784, the Key
Project of Chinese Ministry of Education (No. 205110), the Youth
Scientific Research Fund of Hunan Provincial Education Department
under Grant No. 07B085 and the Foundation of CSUFT under Grant No.
06Y020.
\end{acknowledgments}

\begin{figure}[htbp]
\includegraphics[width=9cm]{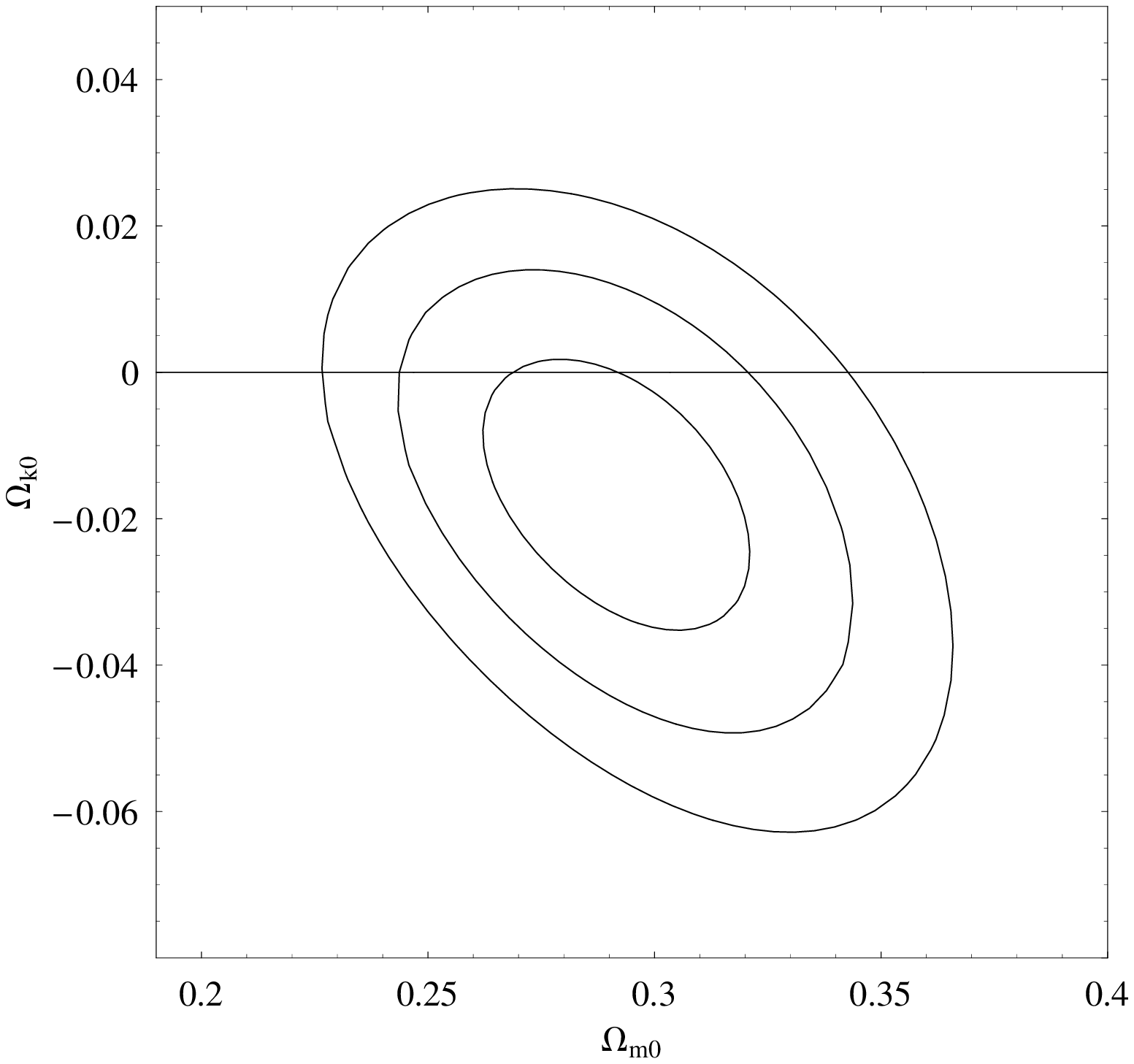}
\caption{\label{Fig1} The $1\sigma$, $2\sigma$ and $3\sigma$
confidence contours for a $\Lambda CDM$ universe with  curvature
from the combination of Sne Ia, BAO, CMB, the X-ray gas mass
fraction in clusters and Hubble parameter data.}
\end{figure}

\begin{figure}[htbp]
\includegraphics[width=5cm]{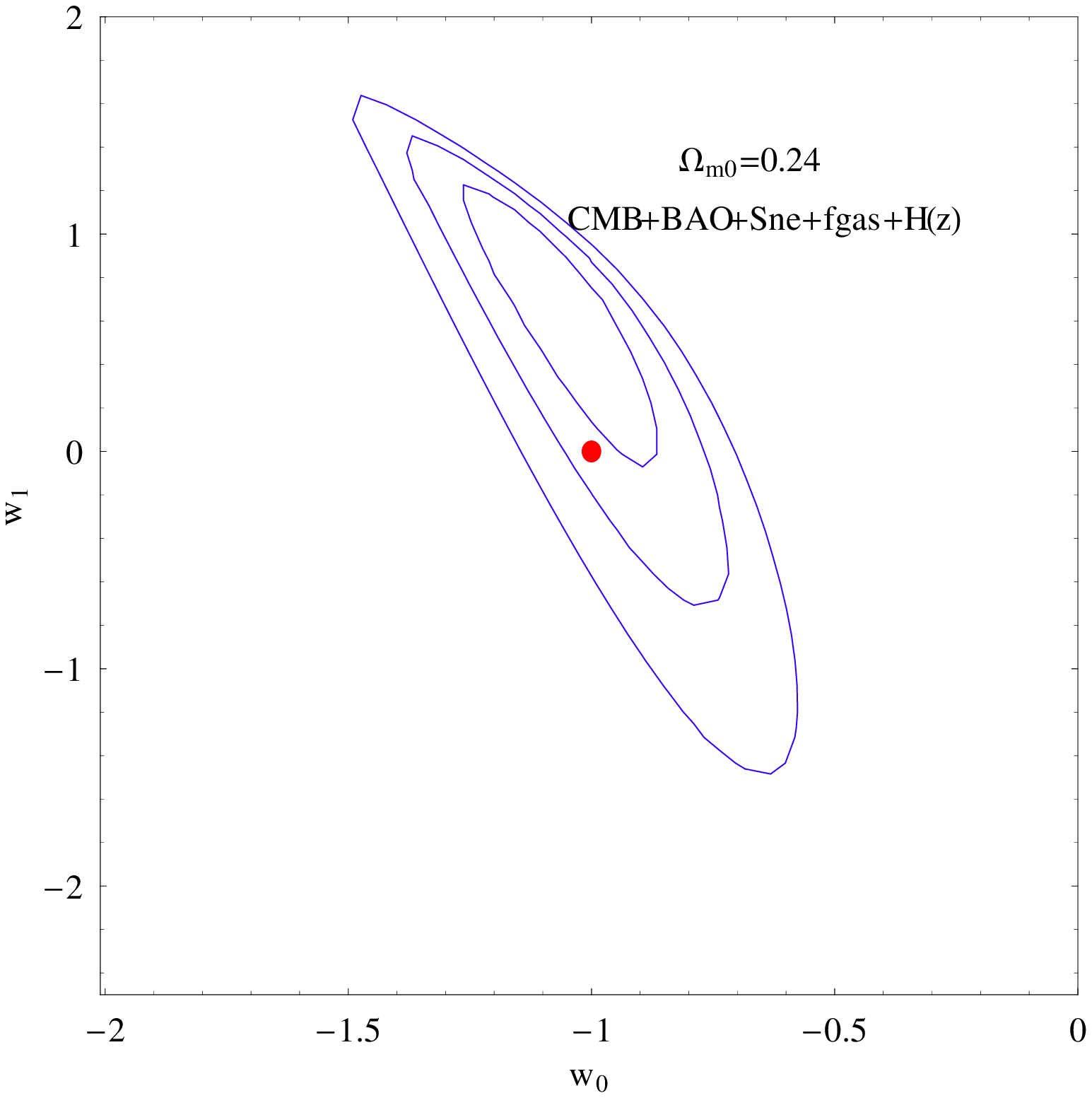}\includegraphics[width=5cm]{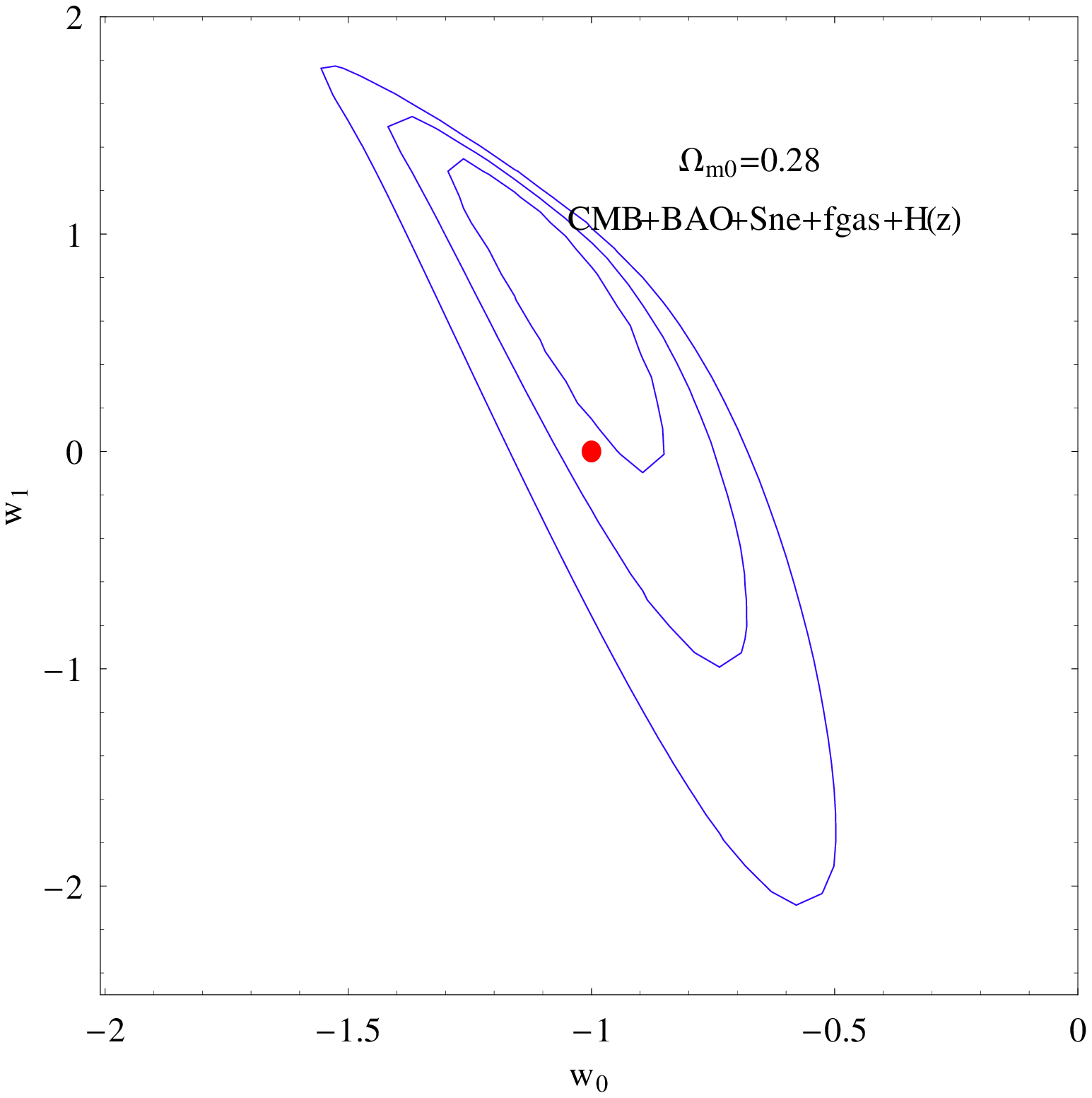}\includegraphics[width=5cm]{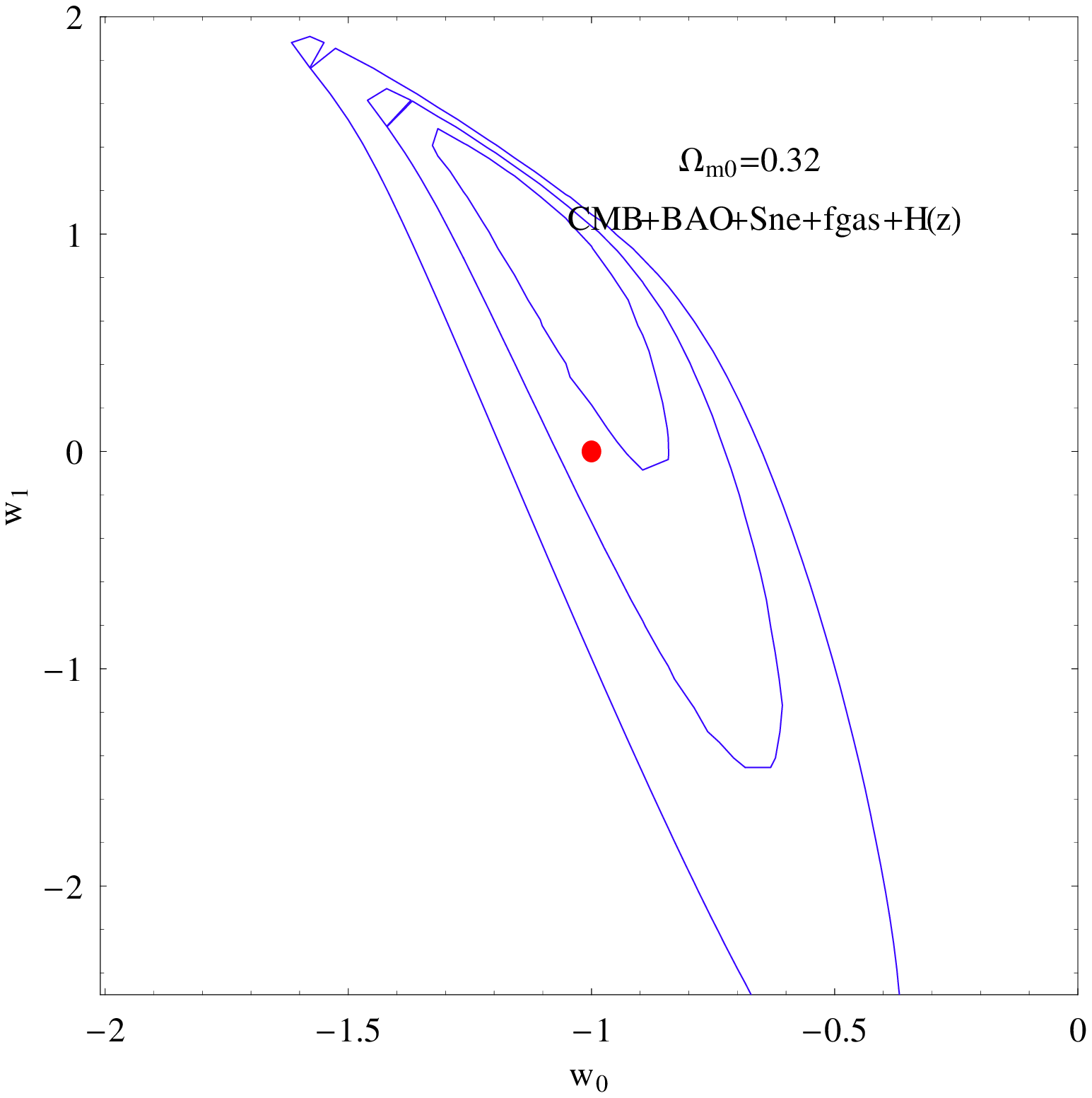}
\includegraphics[width=5cm]{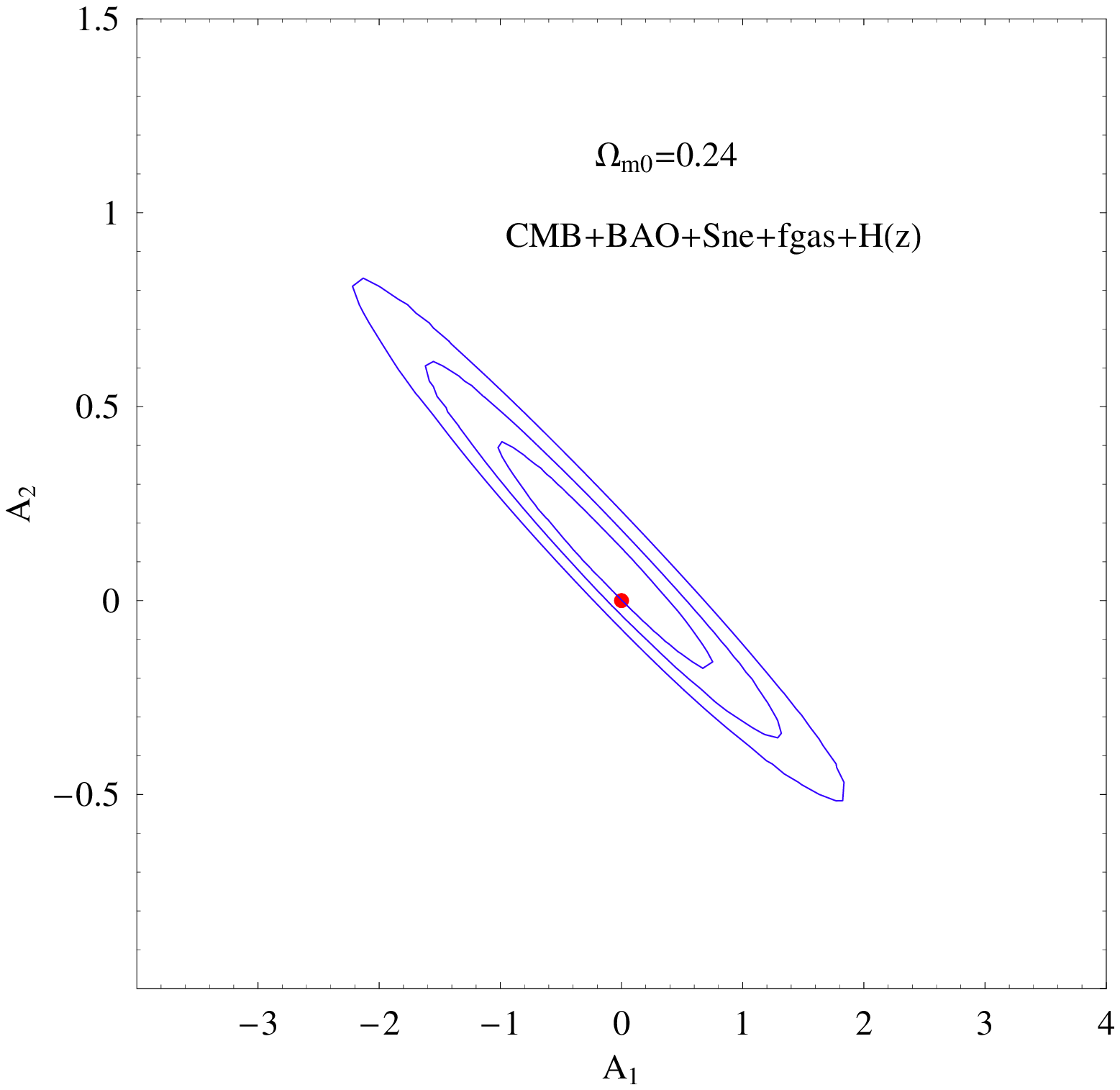}\includegraphics[width=5cm]{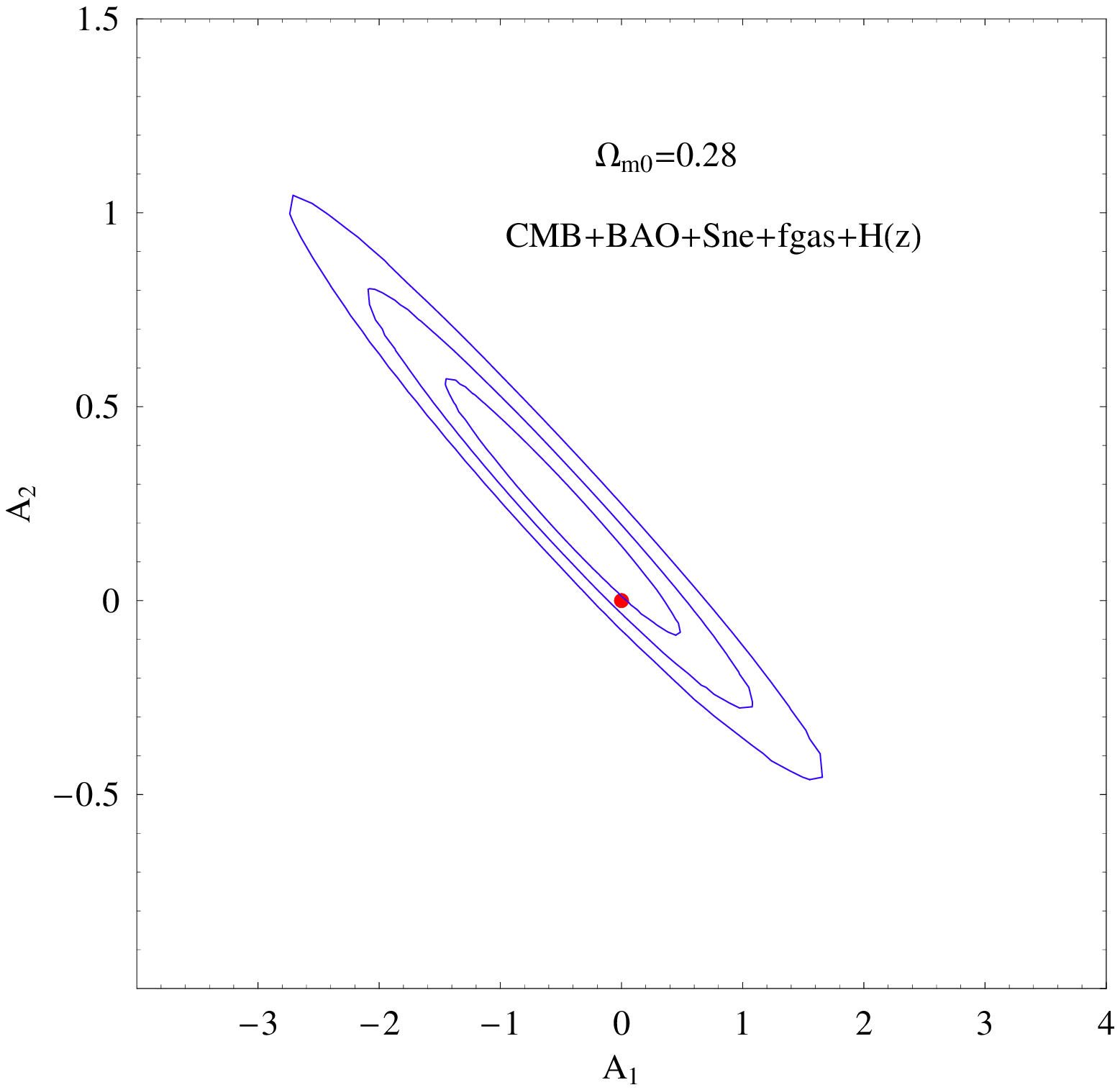}\includegraphics[width=5cm]{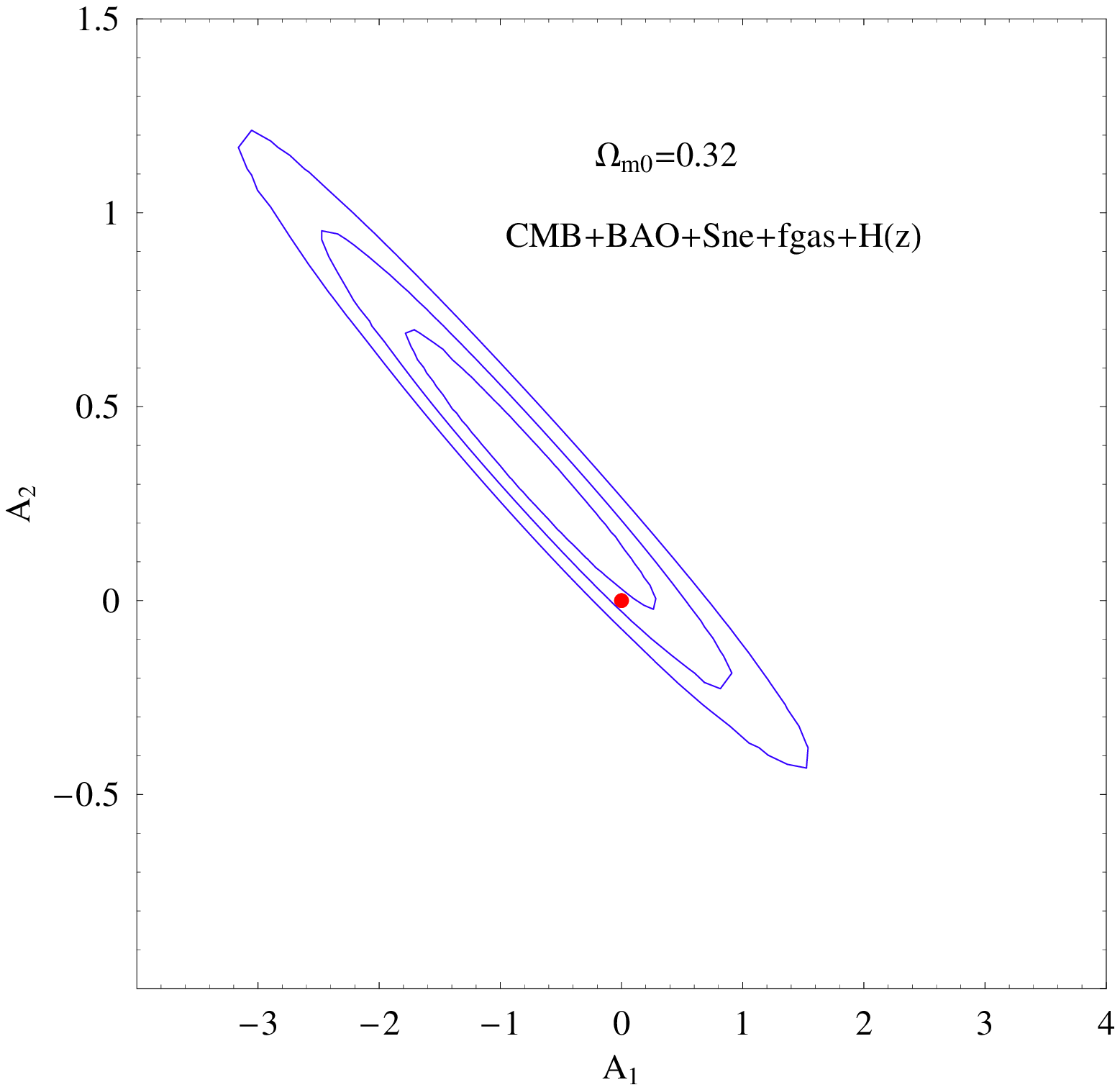}
\caption{\label{Fig2} The constraints on two parameterized dark
energy models from the combination of the Gold Sne Ia,  BAO, CMB,
the X-ray gas mass fraction in clusters and the Hubble parameter
data. The upper and down panels, respectively, show the results of
 $ w(z) = w_0 + w_1 z/(1 +
z)$(Mod1) and
$w(z)=\frac{1+z}{3}\frac{A_1+2A_2(1+z)}{\Omega_{DE}}-1$ (Mod2) with
three different priors over $\Omega_{m0}$: $\Omega_{m0}=0.24$,
$\Omega_{m0}=0.28$ and $\Omega_{m0}=0.32$. The red dot represents
the $\Lambda CDM$ model. }
\end{figure}

\begin{figure}[htbp]
\includegraphics[width=5cm]{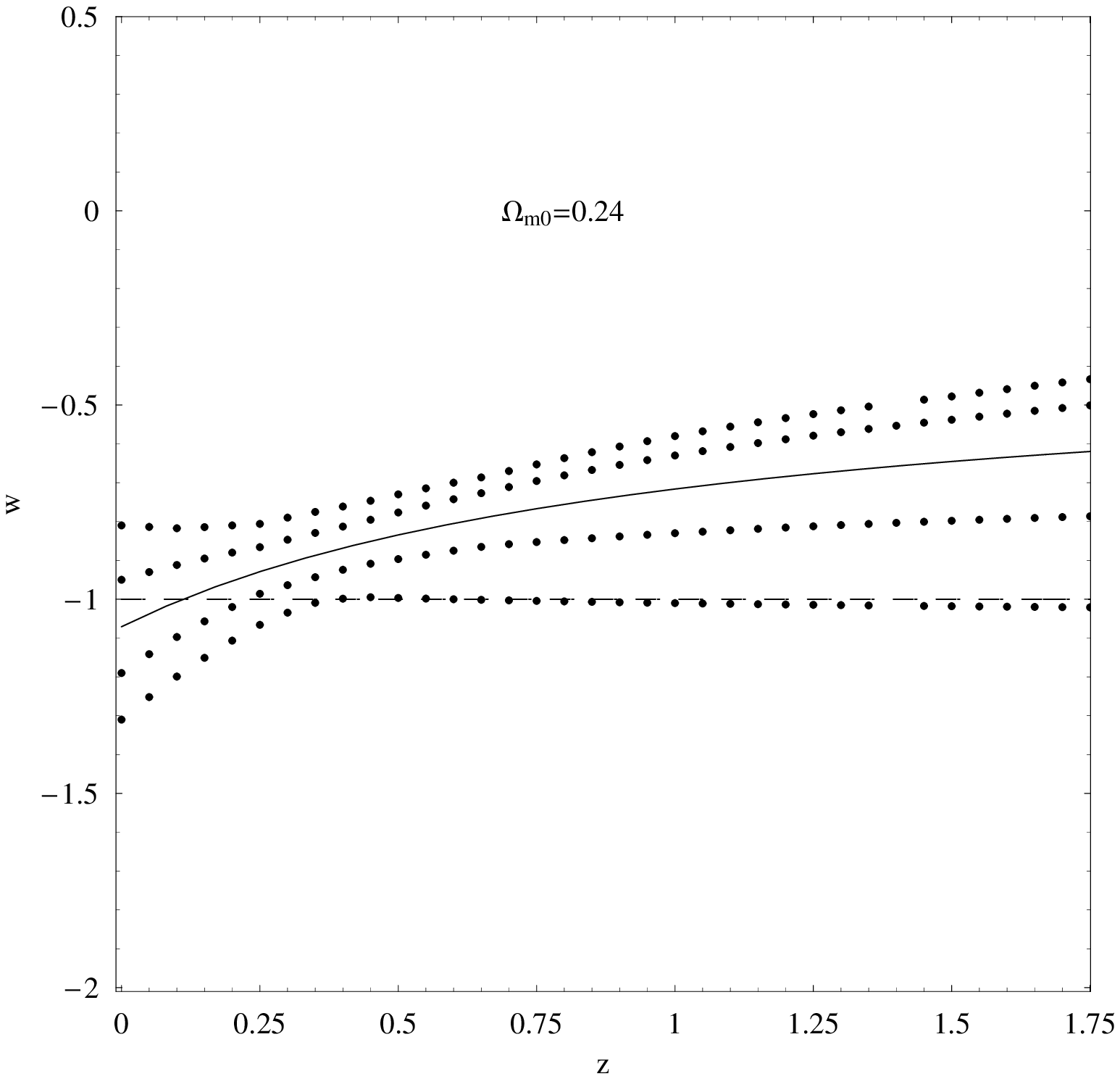}\includegraphics[width=5cm]{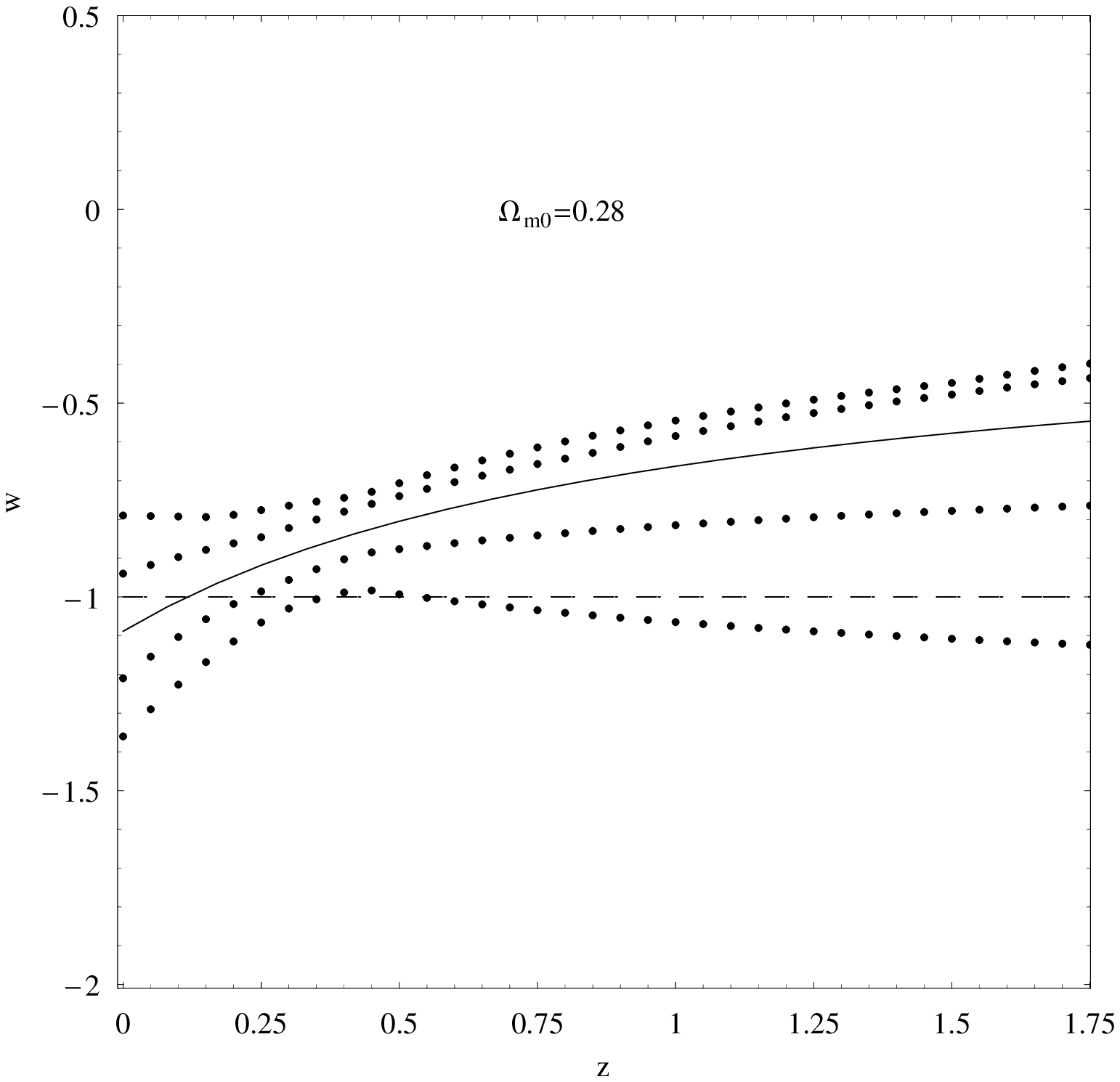}\includegraphics[width=5cm]{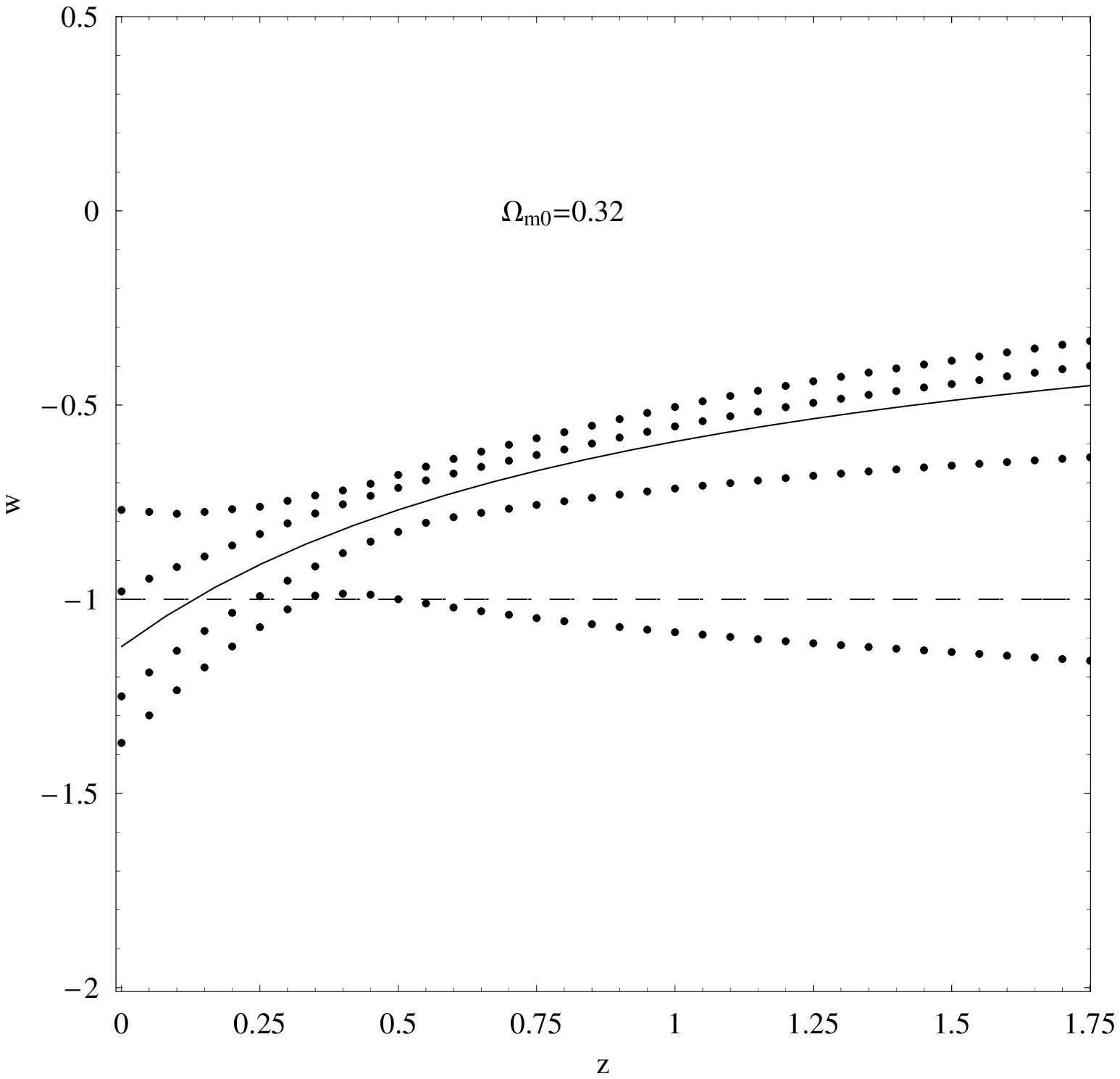}
\includegraphics[width=5cm]{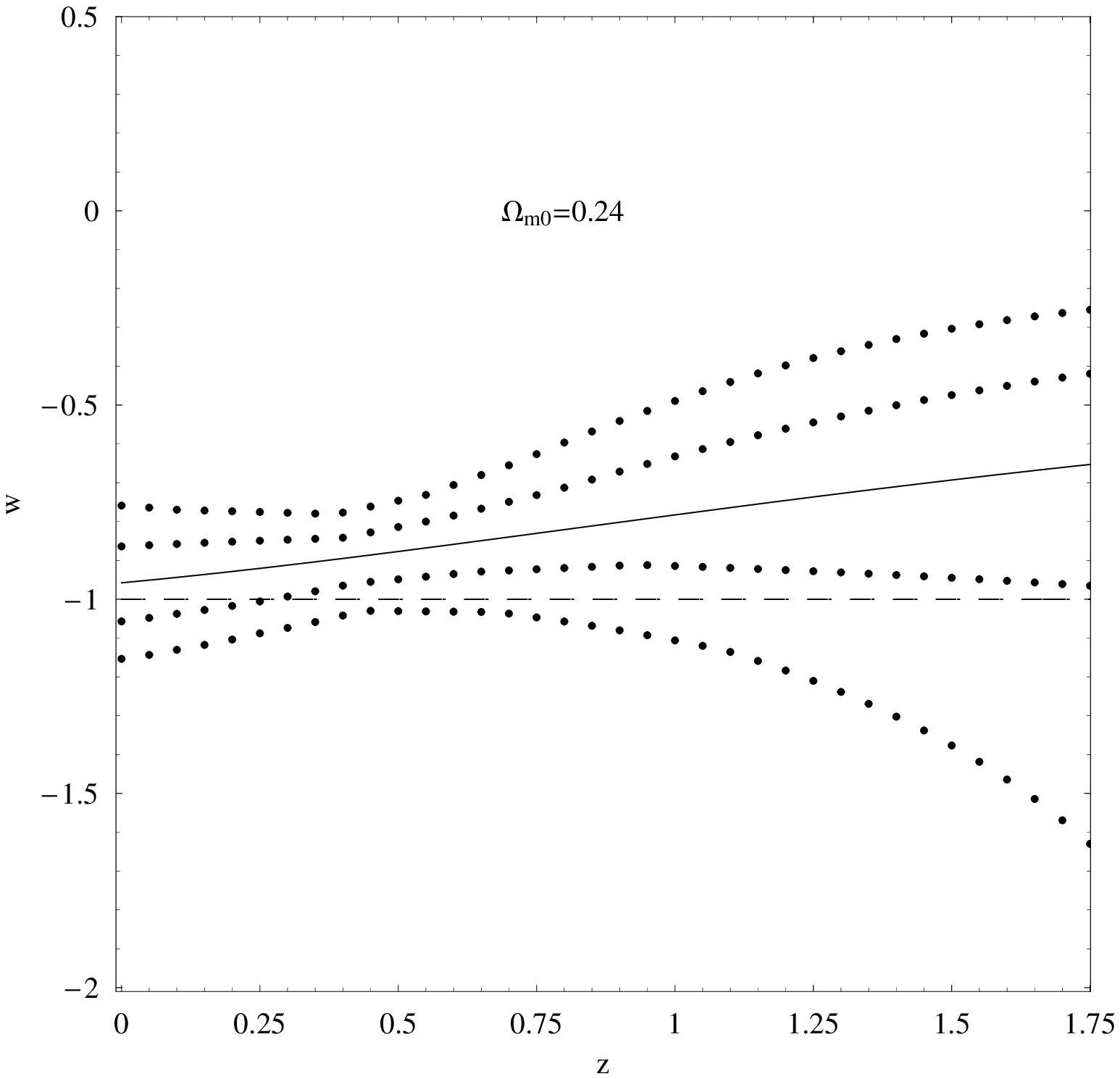}\includegraphics[width=5cm]{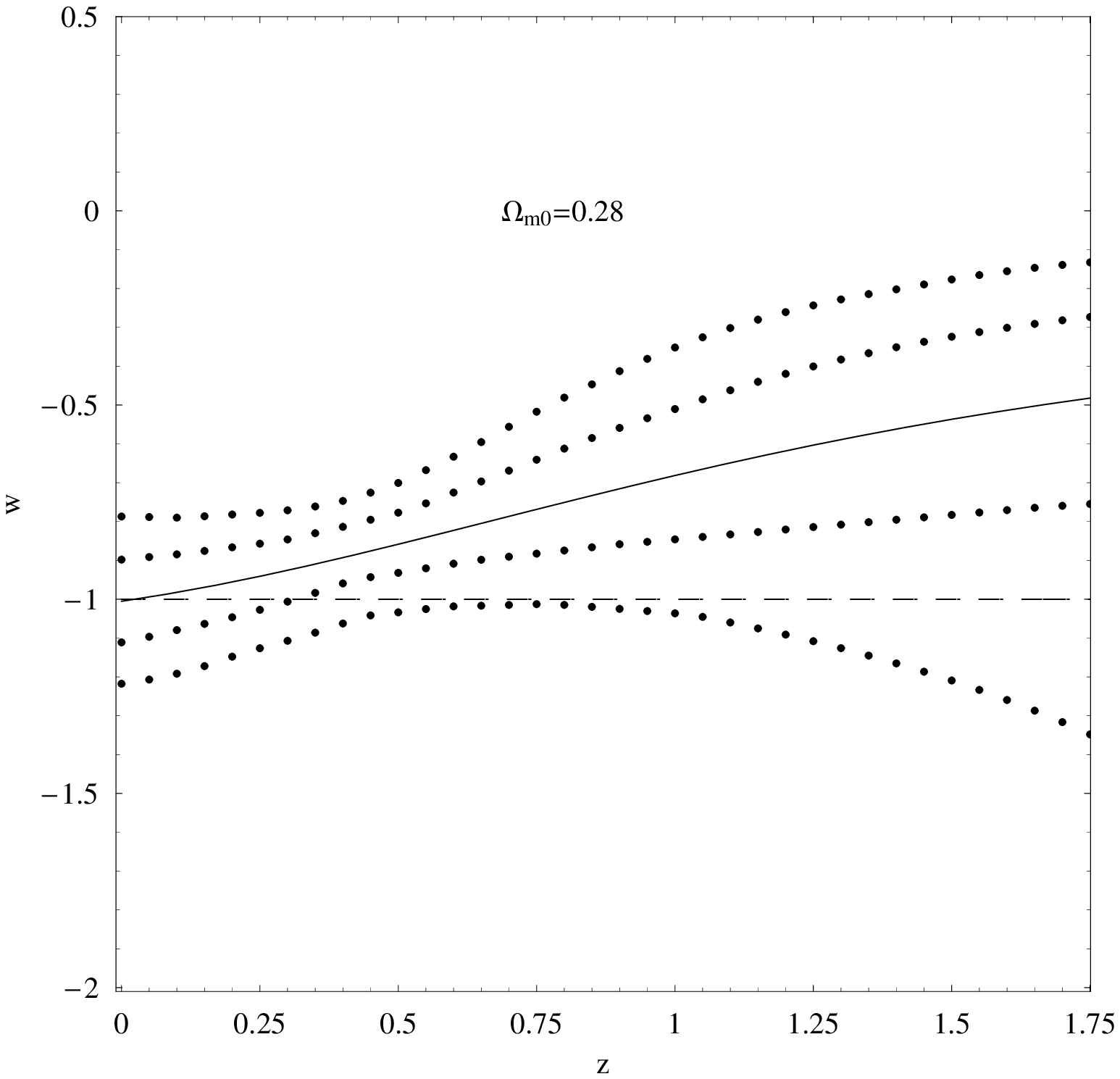}\includegraphics[width=5cm]{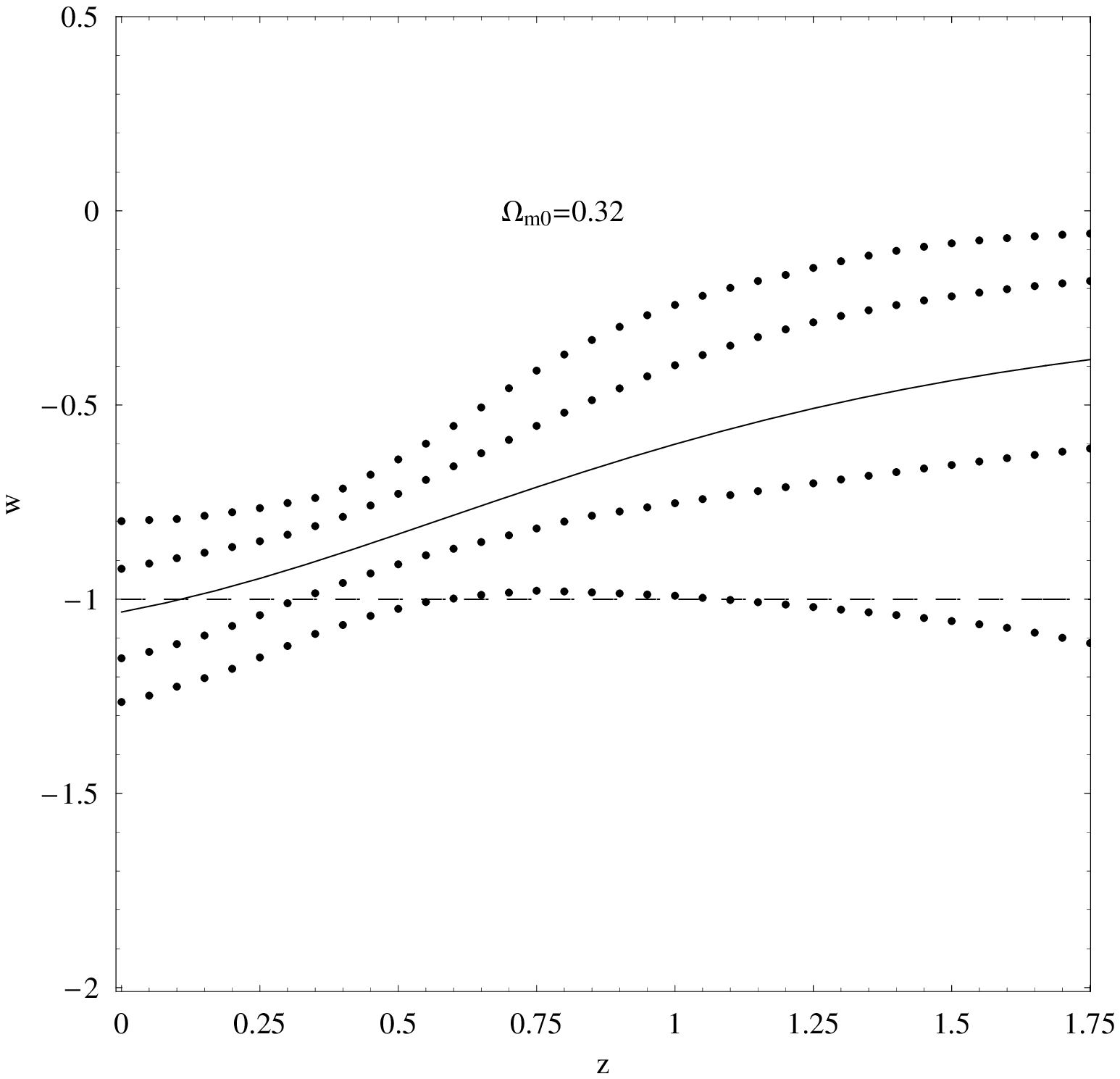}
\caption{\label{Fig3}The evolution of $w(z)$. The upper and down
panels, respectively, show the results of
 Mod1 and Mod2 with $\Omega_{m0}=0.24$, $\Omega_{m0}=0.28$ and
$\Omega_{m0}=0.32$. The  solid line shows the evolution of $w(z)$
with the model  parameters at the best fit values, and the dotted
lines are for the $1\sigma$ and $2\sigma$ errors. }
\end{figure}

\begin{figure}[htbp]
\includegraphics[width=5cm]{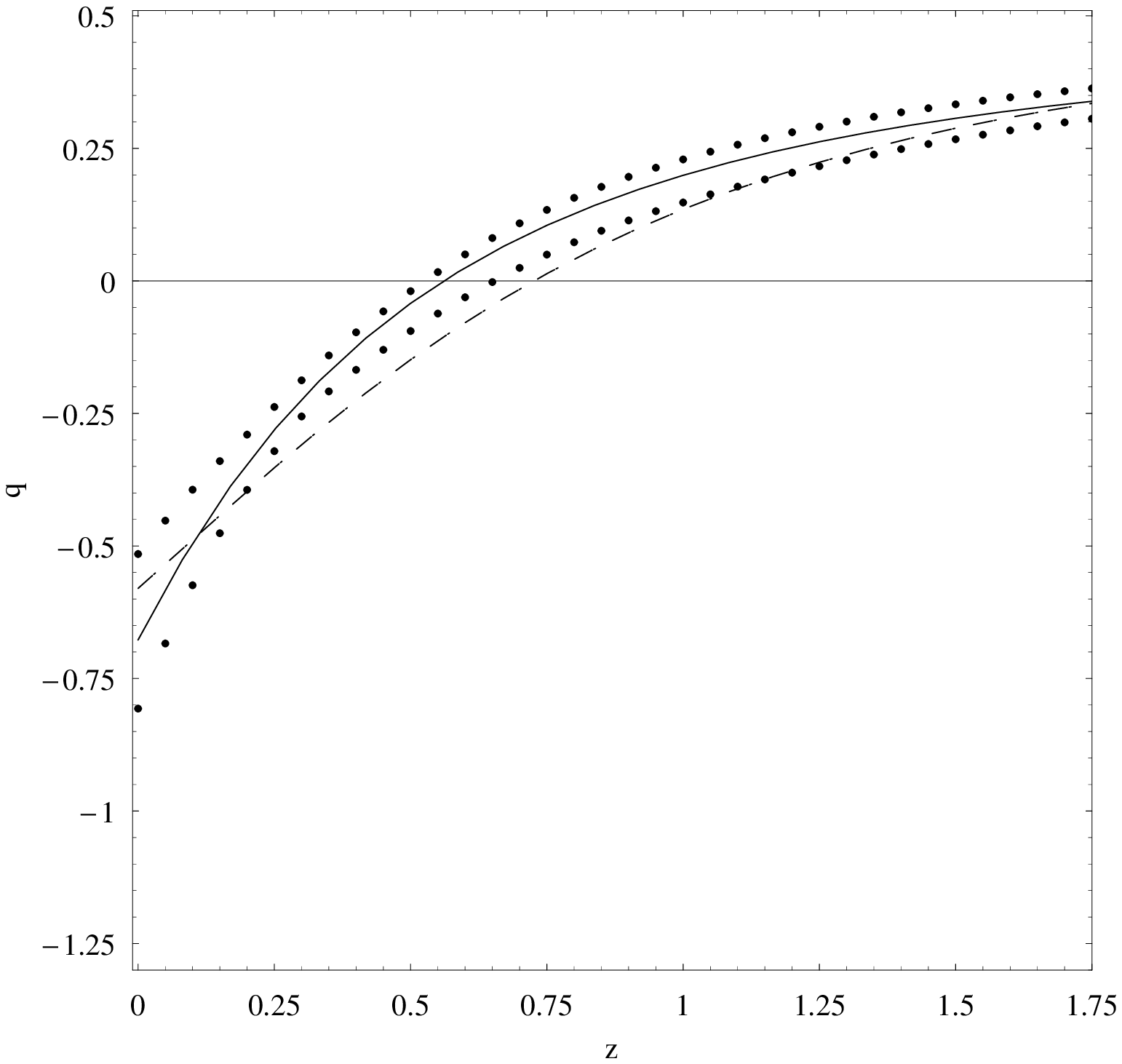}
\includegraphics[width=5cm]{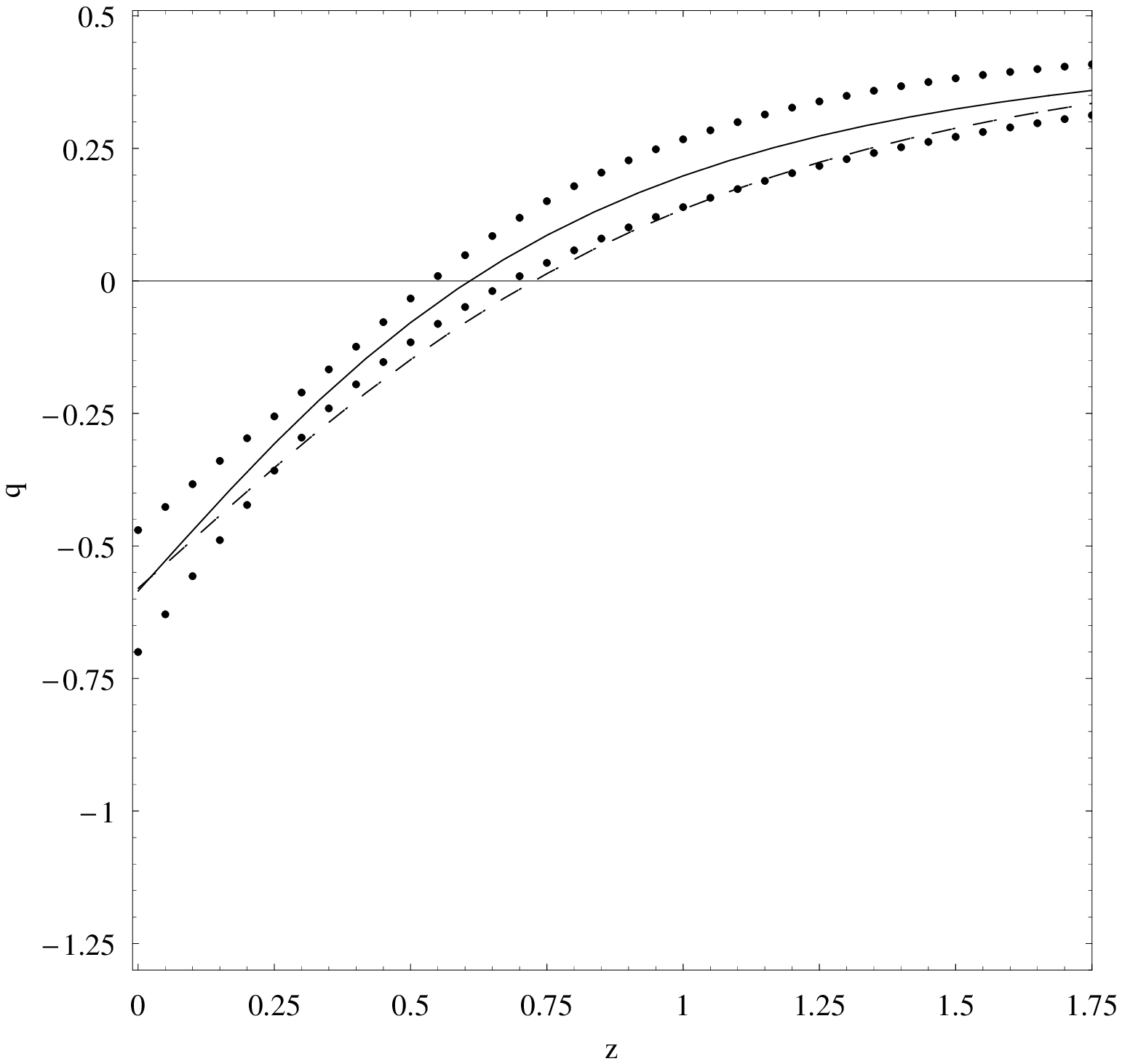}
\caption{\label{Fig4}The behavior of $q(z)$. The left and right
panel, respectively, shows the results of Mod1 and Mod2 with
$\Omega_{m0}=0.28$. The dashed line represents the $\Lambda CDM$
with $\Omega_{m0}=0.28$, the solid line shows the evolution of
$q(z)$ with $\Omega_{m0}=0.28$ and the model parameters at the best
fit values, and the dotted lines are for the $1\sigma$ error. }
\end{figure}

\end{document}